\documentclass[prd,aps,twocolumn,showpacs,,preprintnumbers,amsmath,amssymb]{revtex4}
\usepackage[dvips]{graphicx}
\usepackage{amsmath}
\usepackage{amssymb}

\usepackage{graphicx}
\usepackage{dcolumn}
\usepackage{bm}
\def\be{\begin{equation}}
\def\ee{\end{equation}}
\def\bea{\begin{eqnarray}}
\def\eea{\end{eqnarray}}

\begin{document}


\date{\today}

\title{Is the Spectrum of Gravitational Waves the ``Holy Grail'' of Inflation?}

\author{Robert H. Brandenberger} \email[email: ]{rhb@hep.physics.mcgill.ca}

\affiliation{Department of Physics, McGill University,
Montr\'eal, QC, H3A 2T8, Canada}


\begin{abstract}

It is often said that detecting a spectrum of primordial gravitational waves 
via observing B-mode polarization of the Cosmic Microwave Background is
the ``Holy Grail'' of inflation. The purpose of this short note is to point
out that it is indeed of immense scientific interest to search for a signal
of gravitational waves in B-mode polarization. However, rather than proving
that inflation is the right paradigm of early universe cosmology, a positive 
signal of direct primordial B-mode polarization might well be due to other
sources than inflation. In fact, a careful characterization of the spectrum
of B-mode polarization might even falsify the inflationary paradigm.

\end{abstract}

\maketitle

\newcommand{\eq}[2]{\begin{equation}\label{#1}{#2}\end{equation}}

\section{Overview}

At the present time, the Cosmic Microwave Background (CMB) 
is our most accurate tool to study the structure of the universe.
CMB temperature maps allow us to study the structure of the 
universe at the time $t_{rec}$ when the microwave photons last
scattered. The current high precision maps provide strong
support for the paradigm that the structure which we now see
in our universe comes from a primordial spectrum of almost
adiabatic and almost scale-invariant fluctuations which were
present on super-Hubble scales already before $t_{rec}$. 
The detailed quantitative analysis of these maps also allows us
to tightly constrain a number of cosmological parameters which
describe our background cosmology. The CMB temperature maps
are now so accurate that they are cosmic variance limited for
all larger angular scales. 

There is, however, more information in the CMB than simple
temperature maps reveal: CMB radiation is polarized, and the
polarization carries a lot of important information about
cosmology. CMB polarization can be decomposed into
E-mode and B-mode polarization. The E-mode polarization
power spectrum has now been observed, as has the cross correlation
power spectrum between temperature and E-mode. However,
to date no B-mode polarization has been detected.

Primordial adiabatic density fluctuations only lead to E-mode
polarization, whereas gravity waves will lead to a signal in
B-mode polarization. Thus, the search for B-mode
CMB polarization is one of the ways to detect a spectrum of
primordial gravitational radiation on cosmological scales.

The inflationary scenario \cite{Guth, others} is the current paradigm for
early universe cosmology. Inflation solves some of the mysteries
of Big Bang Cosmology and is the first model based on
principles of relativistic physics to predict \cite{Mukh} (see also
\cite{Starob2, others2}) an approximately
scale-invariant spectrum of approximately adiabatic cosmological
fluctuations \footnote{The word ``approximately" is to be understood
slightly differently in the two contexts above. All simple inflationary
models lead to a slight deviation from scale-invariance - this
explains the meaning of ``approximate" in the first case. The simplese
inflationary models produce a purely adiabatic spectrum of
fluctuations, but a generic more complicated model will produce an
iso-curvature component to the fluctuation spectrum - this is the
way in which the second use of ``approximate" should be understood.}.
Inflationary models also predict a roughly scale-invariant
spectrum of graviational waves \cite{Starob2}. 

It is currently often said that the measurement of gravitational radiation
through the detection of primordial B-mode polarization is the ``Holy
Grail" needed to confirm that inflation is the correct theory of the
very early universe (see e.g. \cite{White}). The purpose of this note is to put 
this claim into
some perspective. Implicit in the claim that detecting primordial B-mode
polarization will be the ``Holy Grail" of inflation are two assumptions:
firstly that there are no other sources of direct B-mode polarization except
for gravitational waves, and secondly that if indeed gravitational waves
are responsible for B-mode polarization, then these waves will come from
inflation. 

Here, I point out that IF there is a measurable spectrum of gravitational waves
on cosmological scales, these waves could well be due to other sources
than inflation. Whereas inflation typically produces a small amplitude of
gravitational waves, the other sources which I will mention can produce
a larger amplitude. Secondly, I point out that there are other cosmological
sources of B-mode polarization. Hence, a positive detection of B-mode
polarization might NOT be due to gravitational waves at all.

This Note is intended mainly for experimentalists. I hope to argue here
that it is even more important to search for B-mode polarization than
if gravitational waves from inflation were the only possible source. In
fact, I will show that it is possible that B-mode polarization results could
falsify the inflationary paradigm. 

\section{Why Look Beyond Inflationary Cosmology?}

The scenario of cosmological inflation  \cite{Guth} has for good reasons
become the paradigm of early universe cosmology. It explains mysteries
about the structure of the universe which Standard Cosmology could
not address such as the horizon, flatness and entropy problems.
More importantly, it provides a theory for the origin of an almost
scale-invariant spectrum of primordial adiabatic curvature fluctuations
on scales which at $t_{rec}$ are larger than the Hubble radius (the
apparent horizon). Such a spectrum is in excellent agreement with the
current data.

\begin{figure} 
\includegraphics[height=8.5cm]{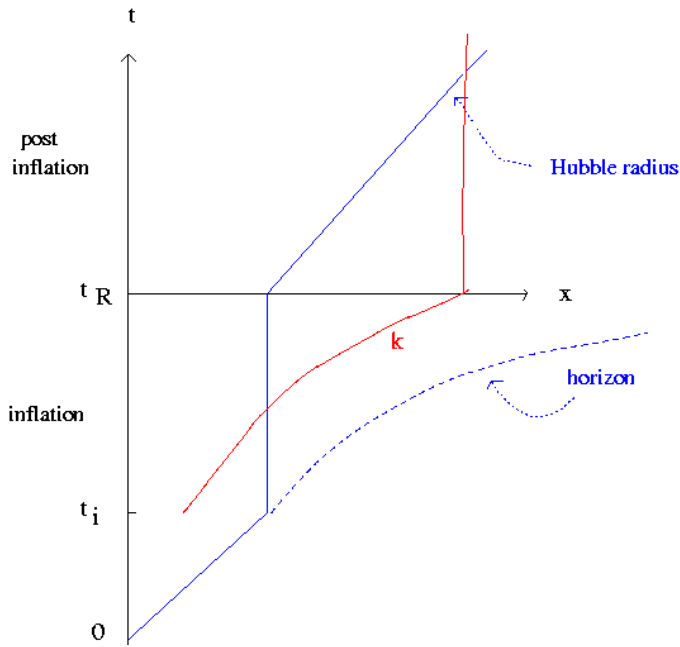}
\caption{Space-time sketch of inflationary cosmology.
The vertical axis is time, the horizontal axis corresponds
to physical distance. The period of inflation lasts from
$t_i$ until $t_R$. Before $t_i$ and after $t_R$ the
universe is in a phase dominated by radiation. The
time $t_R$ can be viewed as the start of the the
Standard Cosmology epoch of the inflationary scenario.
The solid line labelled $k$ is the
physical distance between two points at rest in the
expanding space. This distance corresponds to the
scale of a fluctuation. Note that the horizon (the forward
light cone of a point at the Big Bang) begins to deviate
exponentially from the Hubble radius as the inflationary
phase proceeds.}
\label{infl1}
\end{figure}

A space-time sketch of inflationary cosmology is shown in Figure 1.
The vertical axis is time, the horizontal axis represents physical
spatial distance. The inflationary phase lasts from time $t_i$ until
the time $t_R$ of ``reheating". During this phase, space is expanding
almost exponentially and hence the Hubble radius (defined as the
inverse expansion rate) 
is approximately constant \footnote{The Hubble radius separates
small scales where fluctuations typically oscillate from large
scales where the microphysical oscillations are frozen out and the
evolution of the perturbations are dominated by the gravitational
dynamics of the background. See \cite{MFB} for an in-depth treatment
of the theory of cosmological perturbations, and \cite{RHBrev} for
an introductory overview.} During the inflationary phase, the
physical wavelength of a fluctuation mode increases exponentially.
It is this expansion which leads to the possibility that the fluctuation
modes on large cosmological scales which are currently being
probed by observation emerge at the beginning of the inflationary
period on sub-Hubble scales, which allows a causal theory for
their origin. In inflationary cosmology it is assumed that the
fluctuations emerge at some early time as quantum vacuum
perturbations on sub-Hubble scales \cite{Mukh} (see also
\cite{Starob2, others2}). The success of inflation as a theory
that can explain the origin of structure in the universe is based
on two key points: Firstly, scales which we observe today need
to have a wavelength smaller than the Hubble radius at some
initial time. Secondly, the fluctuations need to propagate freely
for a long duration of time on super-Hubble scales. This is
necessary to establish the coherence of the phases of different
fluctuation modes which in turn is responsible for producing
the acoustic oscillations in the angular power spectrum of the
CMB. Any alternative to inflation will have to contain these two
features. 

It is important to realize that a scale-invariant spectrum of adiabatic
fluctuations as the origin of structure in the universe was already
postulated a decade before the development of inflationary cosmology
\cite{early}, and that ANY scenario which produces an almost
scale-invariant spectrum of adiabatic fluctuations agrees equally
well with the observations. This point is generally recognized, but
it is then often claimed that since inflation produces a nearly
scale-invariant spectrum of gravitational waves, the discovery
of such a spectrum will confirm inflation. However, as will be
emphasized here, this logic is incorrect.

Part of the motivation for looking beyond inflationary cosmology is
the fact that (at least current realizations of) inflation suffer from
various conceptual problems (see e.g. \cite{problems} for more
extended discussions of these problems). Current models of
inflation are based on coupling scalar matter fields to Einstein
gravity. Whereas it is easy to construct toy models of inflation,
it has proven very difficult - in spite of thirty years of efforts - to
embed inflation into a fundamental theory of matter (see e.g.
\cite{stringinflation} for reviews of attempts to embed inflation into
superstring theory).  Even if a candidate model with a scalar field
which can lead to inflation is found, then often special initial
conditions need to be chosen in order to obtain inflation, and
the parameters in the model need to be finely tuned if the observed
amplitude of the spectrum of fluctuations is to emerge (this is
the so-called ``amplitude problem" of inflationary cosmology).

There are, however, more serious problems. One of them is the
``trans-Planckian problem" for fluctuations \cite{RHBrev0,Jerome}: if the
period of inflation lasts only slightly longer than it needs to in
order for inflation to be successful at solving the cosmological problems of
the Standard Big Bang which it is designed to solve \footnote{In numbers, 
more than $70 H^{-1}$ (where $H$ is the Hubble expansion rate during
the inflationary phase), while inflation needs to last at least $50 H^{-1}$ -
there is a slight dependence of these numbers on the energy scale at
which inflation takes place - we have assumed the standard value.},
then the wavelengths of all scales which are currently probed in 
cosmological observations are smaller than the Planck length at the
beginning of the inflationary period. The physical laws which underlie
inflationary cosmology break down on length scales smaller than the
Planck length (possibly much earlier), and thus the usual computations
of the spectrum of cosmological fluctuations have been done using
a theory which does not apply in the zone of ignorance indicated in
Figure 2. It is easy to construct \cite{Jerome} toy models of physics
beyond the Planck scale which yield large deviations from the
usual predictions of inflation.

\begin{figure} 
\includegraphics[scale=0.5]{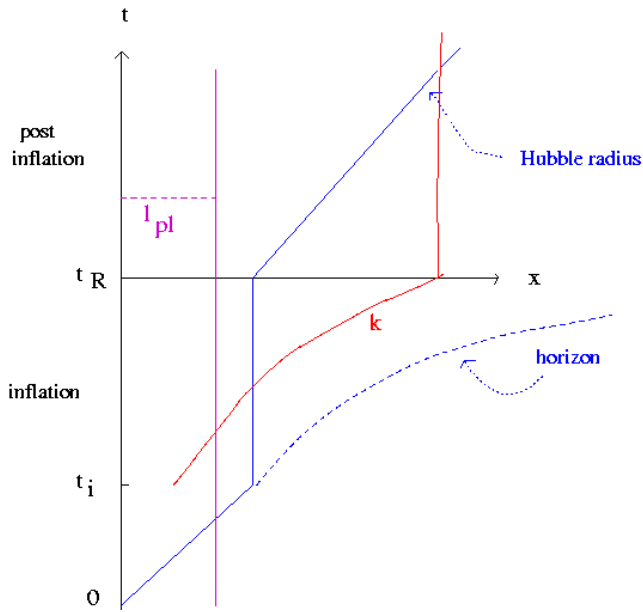}
\caption{A space-time sketch of inflationary cosmology. As in Figure 1, the
vertical axis is time, the horizontal axis denotes physical distance. In
addition to the Hubble radius, the horizon and the physical wavelength
corresponding to a fluctuation mode which expands as the size of space
increases (the curve denoted by $k$), we show the vertical curve coresponding
to the Planck length. Unless the period of inflation is only very slightly larger
than the minimal duration it has to have  to solve the problems of Standard Cosmology,
the physical wavelength corresponding to our current Hubble radius is smaller
than the Planck length at the beginning of inflation. Thus, all scales which are
observable now in cosmology emerge from the short wavelength zone of ignorance.
This is an illustration of the trans-Planckian problem.}
\end{figure}

If inflation is obtained by coupling classical matter fields to Einstein
gravity, then a cosmological singularity in the past is as unavoidable
\cite{Borde} as it is in Standard Cosmology. Thus, in the same way 
that we knew that Standard Cosmology could not be the ultimate
theory of the very early universe, we know that neither can inflationary
cosmology. There must be a theory beyond inflation which applies
at very high densities. It may be the case that this theory contains
inflation as the low density limit, but this need not be the case. 

In large field toy models of inflation, the energy scale at which the 
inflationary phase
takes place is typically of the order $10^{16} {\rm GeV}$, much closer
to the Planck scale than to scales where the theory has been tested.
In all approaches to quantum gravity it emerges that the Einstein
action does not yield a good approximation to the dynamics at scales
that approach the Planck scale. Since the scale of inflation is so close
to the Planck scale (and even closer to the string scale \cite{GSW}),
it may be dangerous to trust the conclusions obtained by applying
Einstein gravity.

The ``Achilles heel" of inflation is related to the cosmological constant
problem which arises whenever we couple quantum matter to
gravity. Quantum vacuum energy cannot gravitate since if it would
it would lead to a cosmological constant which is at least sixty
orders of magnitude  larger than the observed one. Thus, there must
be a mechanism which renders quantum vacuum energy
gravitationally inert. Why doesn't the same mechanism render
the constant part of a scalar field potential energy (which leads
to inflation) gravitationally inert? 

The above conceptual problems of inflation motivate the search
for alternatives. As will be shown below, there are in fact alternative
cosmological scenarios, and some of them yield an almost
scale-invariant spectrum of gravitational waves like inflationary
cosmology does. In fact, in at least one of these alternative
approaches, the amplitude of the gravitational wave spectrum is
larger than it is in simple inflationary models.

\section{Alternatives to Inflation}

In this section I will briefly review three alternative scenarios to
inflation. This list is not meant to be complete! The three
scenarios I will mention are an ``emergent stringy universe"
\cite{BV,NBV}, the ``matter bounce scenario" (see e.g.
\cite{RHBrev3} for a review of this and other alternatives),
and the Ekpyrotic universe \cite{Ekp}. Two of the three
produce a spectrum of almost scale-invariant gravitational
waves on cosmological scales, both with amplitudes comparable
or larger than what small field inflationary models produce.

\subsection{Emergent Stringy Universe}

\begin{figure} 
\includegraphics[scale=0.5]{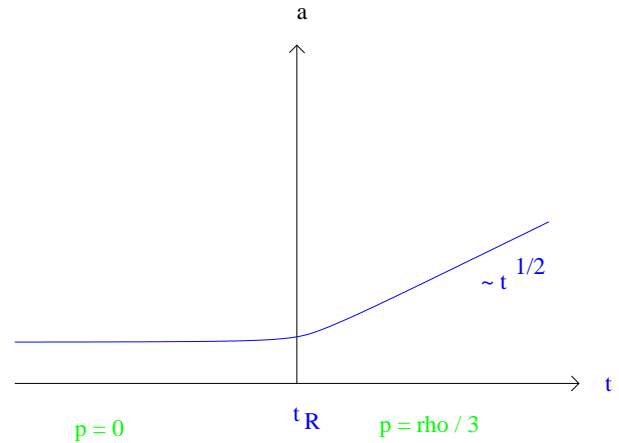}
\caption{Time evolution of the scale factor (size of space) in the emergent universe
scenario. The time axis is the horizontal axis, the vertical axis gives the value of
the scale factor. The time $t_R$ is the end of the emergent phase, the time when
the universe makes the transition to the radiation stage of Standard Cosmology
during which the scale factor increases as the square root of time.}
\end{figure}

In the ``emergent universe" scenario \cite{emergent} it is
postulated that the universe begins in a long quasi-static
phase which at some point in time $t_R$ makes a transition to the
expanding universe of Standard Big Bang Cosmology. The
evolution of the scale factor as a function of time is
sketched in Figure 3, the horizontal axis being time, the
vertical axis indicating the value of the scale factor. A
sketch of the resulting space-time is given in Figure 4.
As in Figures 1 and 2, the vertical axis is time and the
horizontal axis indicates physical distance. We plot the
Hubble radius and the wavelength of some fluctuation mode.
Since the universe is quasi-static before $t_R$, the Hubble
radius tends to infinity. Hence, fluctuations which we observe
today originate on sub-Hubble scales during the early
``emergent" phase. After $t_R$, the fluctuations propagate
on super-Hubble scales until they re-enter the Hubble
radius at late times. Thus, the two key properties allowing
for the existence of a causal mechanism to produce primordial
cosmological fluctuations and which also lead to acoustic
oscillations in the CMB angular power spectrum (see the
discussion at the end of the second paragraph of Section 2)
are satisfied.

\begin{figure} 
\includegraphics[scale=0.4]{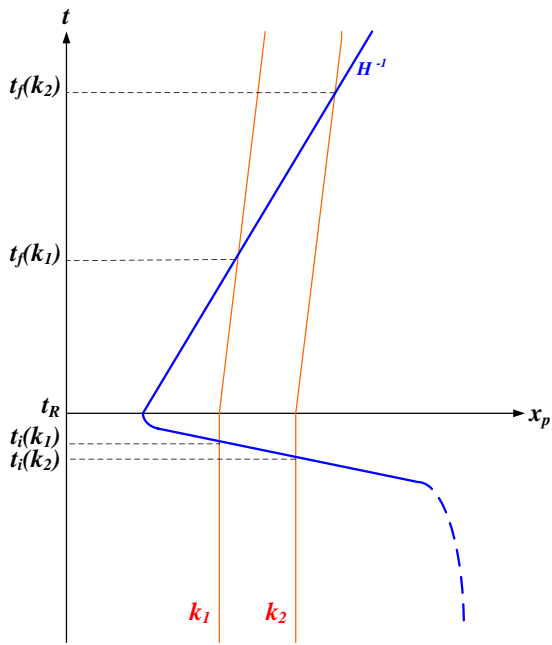}
\caption{Space-time sketch of the emergent universe scenario. As in Figure 1
in the illustration of inflationary cosmology, the vertical axis indicates time, the
horizontal axis physical distance. The time $t_R$ is the end of the emergent
phase and the onset of the radiation phase and plays a similar role to the
time of reheating $t_R$ in inflationary cosmology. The Hubble radius (the blue
solid curve labelled by $H^{-1}$) is infinite deep in the emergent phase. The
physical wavelengths of fluctuation modes (indicated by $k_1$ and $k_2$ in the
graph) are constant in the emergent phase. A mode exits the Hubble radius towards
the end of the emergent phase at the time $t_i(k)$ and re-enters in late time
cosmology at the time $t_f(k)$. }
\end{figure}

A possible realization of the emergent universe scenario
arises in the context of
``String Gas Cosmology" \cite{BV} (see \cite{SGCrev} for
reviews and \cite{Perlt} for some other original work). If we
couple a gas of closed fundamental strings to a space-time
background in the same way that in Standard Cosmology we
couple a gas of point particles to a background space-time,
then very important differences to point particle cosmology
emerge at high densities. In particular, there is a maximal
temperature for a gas of strings in thermal equilibrium, the
so-called Hagedorn temperature $T_H$ \cite{Hagedorn}. If we
compress a box of strings, then we observe that initially
the temperature $T$ will rise, until $T$ approaches $T_H$.
In the case of a compact space and a gas of closed strings,
the evolution of $T$ as a function of the box radius $R$ is
sketched in Figure 5 \footnote{The decrease of $T$ for
very small values of $R$ is a consquence of the T-duality
symmetry of string theory. Once $R$ becomes smaller than
the string scale, it becomes thermodynamically favorable
for the energy of the string gas to excite string modes which
wind the box, since winding modes become light as $R$
decreases.}. In the string gas cosmology scenario \cite{BV}
it is postulated that the universe begins in a quasi-static
phase where the temperature hovers just below $T_H$.
Eventually, the winding string modes which are important
in the quasi-static phase decay into string loops, and this
leads to the beginning of the radiation phase of Standard
Cosmology.

\begin{figure} 
\includegraphics[scale=0.35]{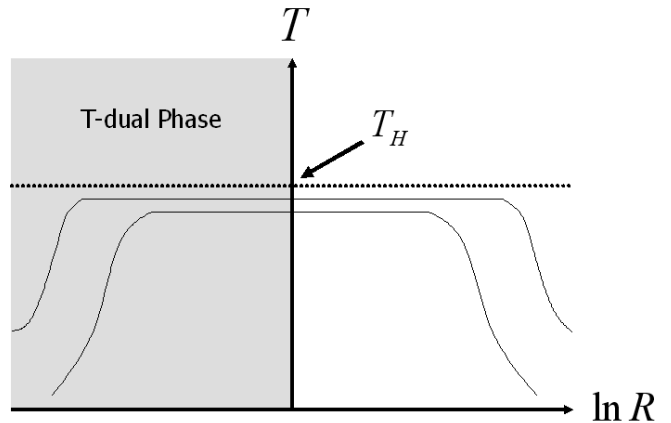}
\caption{Evolution of the temperature (vertical axis) as a function of
the logarithm of the radius (horizontal axis). Two possible evolutionary
tracks are shown which differ in the amount of entropy which they contain.
If we begin the evolution at large radii, then initially the temperature
rises as space contracts as for point particle matter. However, once
the temperature approaches the limiting Hagedorn temperature $T_H$,
the temperature ceases to rise as $R$ decreases. Below a critical radius
which is chosen to be one in the figure, the temperature decreases as
$R$ decreases.}
\end{figure}

In string gas cosmology, matter in the early universe is a gas of
strings in thermal equilibrium. Hence, the thermal fluctuations
dominate over the vacuum ones - in contrast to
inflationary cosmology where it is assumed that fluctuations arise
from a vacuum state. It was realized in \cite{NBV} that in a compact
space, thermal fluctuations of strings
in the early phase of string gas cosmology lead to a
nearly scale-invariant spectrum of cosmological perturbations.
The amplitude of the spectrum depends on the ratio of the
string scale to the Planck scale, and if that number is taken
to be the preferred value in the string theory textbook \cite{GSW},
then the resulting amplitude of the perturbation spectrum is
in good agreement with the data. Thus, string gas cosmology can
successfully address some of the conceptual problems of the
inflationary scenario: there is no amplitude problem for the
magnitude of the spectrum, and since the wavelengths of
fluctuation modes never become smaller than the length they
have at $t_R$ (which is of the order of $1 {\rm mm}$ for
the scale corresponding to the current Hubble radius, and hence
many orders of magnitude larger than the Planck scale), modes
we observe today are never close to the short wavelength zone
of ignorance. Hence the trans-Planckian problem for fluctuations
is absent.

\subsection{Matter Bounce Scenario}

A second alternative to inflation is the ``matter bounce" scenario.
If there were a non-singular bouncing cosmology, then there would
obviously be no singularity problem. Penrose and Hawking
a long time ago taught us that in order to obtain such a 
non-singular bounce, one needs to go beyond General Relativity
as the theory of space-time, or else invoke matter which
violates some of the usual energy conditions. There is
a long history of such attempts (see \cite{Novello} for a
comprehensive list of references). For a recent construction
in the context of a modified gravitational theory (namely 
Ho\v{r}ava-Lifshitz gravity \cite{Horava}) see \cite{RHBHL}
and for another recent construction using ghost condensate
matter see \cite{ghost}.

\begin{figure}[htbp] 
\includegraphics[height=9cm]{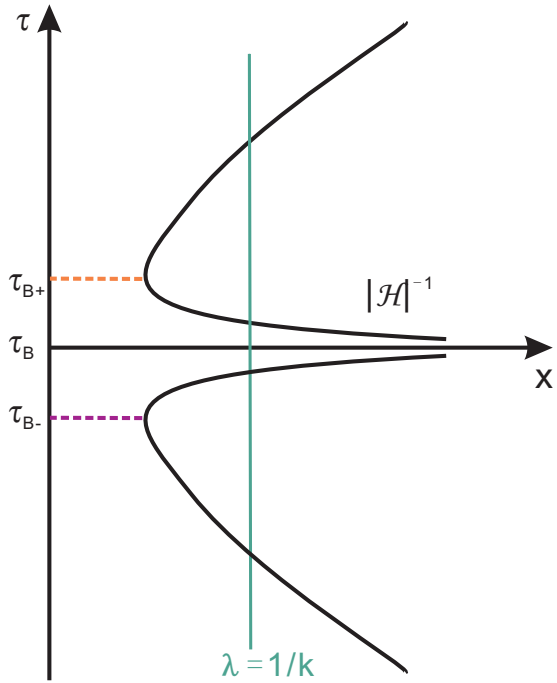}
\caption{Space-time sketch of the matter bounce scenario. The vertical axis
is conformal time $\eta$, the horizontal axis denotes a comoving space coordinate.
Also, ${\cal H}^{-1}$ denotes the co-moving Hubble radius. The
vertical solid green line indicates the wavelength of a particular
perturbation mode. As is evident, this mode exits the Hubble radius
at some point during the contracting phase and re-enters at a later
time in the expanding Standard Big Bang phase.}
\label{bounce}
\end{figure}

Given a non-singular cosmology, it is most logical to assume
that the contracting phase begins with a matter-dominated
phase, then makes a transition to a radiation-dominated
phase before undergoing the bounce and re-emerging
as an expanding Standard cosmology. Figure 6 presents
a space-time sketch of a bouncing cosmology. As in earlier
figures, the vertical axis represents time, the horizontal axis
physical distance. The Hubble radius and the physical wavelength
of perturbation modes are symmetric about the bounce
point. Thus, we see immediately that fluctuations originate
at early times on sub-Hubble scales, and that they
propagate for a long time on super-Hubble scales. Thus,
the two key conditions for being able to generate
primordial curvature fluctuations which agree with the
observed acoustic oscillations in the angular power spectrum
of the CMB are satisfied \footnote{As in the case of string gas cosmology,
this  scenario is free from the trans-Planckian problem for
fluctuations provided that the energy scale of the bounce
is lower than the Planck scale.} . Based on the second
law of thermodynamics we know that the duration of the
radiation phase of contraction cannot be longer than
that of the expanding radiation phase. Hence, scales which are
currently observed have exited the Hubble radius in
the matter-dominated phase of contraction.
  
A while back it was realized  \cite{Wands, Fabio} that initial
vacuum fluctuations on sub-Hubble scales in the contracting
phase evolve into a scale-invariant spectrum of curvature
perturbations on super-Hubble scales before the bounce.
As long as the bounce phase does not change this spectrum
(which it does not in a number of toy models for bouncing
cosmologies in which the evolution of the fluctuations can
be studied (see e.g. \cite{bounceflucts}), we get a scale-invariant
spectrum at late times. This is called the ``matter bounce"
alternative to cosmological inflation as the explanation for
the origin of structure. 

\subsection{The Ekpyrotic Universe}

The major problem of the ``matter bounce" scenario and of other
bouncing cosmologies is the ``anisotropy problem": during the
phase of contraction, the energy density in anisotropies grows
faster than the energy density in the isotropic matter components.
Hence, unless the initial anisotropies are tuned to be extremely
small, no smooth cosmological bounce can occur.

The Ekpyrotic scenario \cite{Ekp} is a bouncing cosmology which avoids
this problem. The model is motivated by a higher dimensional
string theoretical framework \cite{HW} in which our matter is
confined to a four space-time dimensional ``brane" which is one of
the two boundaries of a five dimensional bulk space-time. The
extra spatial dimension is an interval of finite length, bounded on each
side by a brane. There is an attractive force between the two
boundary branes which renders the separation $r$ between the
branes dynamical. From the point of view of our four dimensional
physics on the brane, the separation of the branes corresponds to
a scalar field $\varphi$ with $\varphi \sim {\rm log} r$.

The Ekpyrotic scenario assumes that the attractive potential $V(\varphi)$
is a negative exponential function, leading to slow contraction of
space. Once the separation between the branes becomes comparable
to the string scale, quantum effects take over, leading (as is postulated
but not proven) to a cosmological bounce. The radiation generated
during the bounce leads to a radiation dominated phase of expansion.
In contrast to the ``matter bounce" scenario, the time evolution is not
symmetric about the bounce point! Due to their coupling with $\varphi$,
the scalar cosmological fluctuations acquire a scale-invariant
spectrum in the contracting phase which, when taking extra-dimensional
effects or entropy fluctuations into account (see e.g. \cite{Tolley})
leads to a scale invariant spectrum of curvature fluctuations after
the bounce.

\section{Gravitational Waves from Inflation}

In inflationary cosmology, a scale-invariant spectrum of gravitational
waves is generated from initial quantum vacuum fluctuations of the
canonical fields which represent each polarization state. We begin
with the metric
\be \label{gwansatz}
ds^2 \, = \, a^2(\eta) \bigl[ d\eta^2 - (\delta_{ij} + h_{ij}) dx^i dx^j \bigr] \, ,
\ee
where $h_{ij}$ is a transverse and traceless tensor 
which describes gravitational waves about a cosmological
background given by the scale factor $a(\eta)$. We are using
conformal time $\eta$ which is related to the physical time $t$ via
$dt = a(\eta) d\eta$. The tensor $h_{ij}$ can be decomposed
into the contributions from the two polarization states:
\be
h_{ij}(\eta, x) \, = \, h_{+}(\eta, x) e^{+}_{ij} + h_{-}(\eta, x) e^{-}_{ij} \, ,
\ee
where $e^{+}_{ij}$ and $e^{-}_{ij}$ are the two fixed polarization tensors,
and $h^{+}$ and $h^{-}$ are the two coefficient functions which are
scalar fields on the space-time background.

Inserting the gravitational wave ansatz (\ref{gwansatz}) into the Einstein
action and expanding to quadratic order in $h^{+}$ and $h^{-}$, it is
straightforward to show that the resulting action for the fluctuations takes
canonical form when written in terms of the fields
\be
u^{a} \, \equiv \, a h^{a} \, ,
\ee
where the value of the index $a$ is either $+$ or $-$. The linear
equations of motion for the Fourier modes $u_k$ of $u$ hence
take on the simple form
\be
u_k^{''} + \bigl( k^2 - \frac{a^{''}}{a} \bigr) u_k \, = \, 0 \, ,
\ee
where a prime indicates the derivative with respect to $\eta$.
This equation shows that $u_k$ oscillates on scales smaller than
the Hubble radius but is squeezed on super-Hubble scales
\cite{Grisha}.

Up to this point, the analysis has been completely general.
Let us now consider the application to inflationary cosmology.
In this case, the perturbations originate as quantum
vacuum fluctuations. Since the amplitude of the quantum
vacuum perturbations is given by the Hubble constant $H$, the
power spectrum of gravitational waves at late times is
\be
{\cal P}_h(k) \, \sim \, \frac{H(k)}{m_{pl}} \, ,
\ee
where $H(k)$ is the value of the Hubble constant during the
inflationary phase when the scale $k$ exits the Hubble
radius, and $m_{pl}$ is the Planck mass. 

Note that if inflation is realized in the context of Einstein
gravity and is generated by matter fields obeying the ``Null
Energy Condition", then the Hubble constant must be a
decreasing function of time and hence $H(k)$ decreases as 
$k$ increases. Inflationary cosmology thus produces a
roughly scale-invariant spectrum of gravitational waves with
a tilt which has to be red. This result is to be contrasted with
the tilt of the spectrum of cosmological perturbations produced
by inflation which can be blue (although more often it is red, too).
 
The rate at which the gravitational waves are squeezed on
super-Hubble scales is the same as the rate at which a test
scalar field on a fixed background metric is squeezed. If the
equation of state of the background is independent of time,
the scalar metric fluctuations (the ``cosmological perturbations")
are squeezed at the same rate (this is true e.g. in the matter bounce
scenario during the relevant time intervals), but it is NOT true in
inflationary cosmology where the cosmological perturbations are
amplified more during the reheating period than the gravitational
waves, which leads to the small value of the tensor to scalar
ratio $r$ which inflationary models generally predict:
\be
r(k) \, \equiv \, \frac{{\cal P}_h(k)}{{\cal P}_{\zeta}(k)} \, \ll \, 1 \, ,
\ee
where ${\cal P}_{\zeta}$ is the power spectrum of curvature
fluctuations. Note that large-field inflation models produce a
value of $r$ which is closer to $1$ than small field models.

\section{Alternative Sources of Gravitational Waves}

\subsection{Gravitational Waves from Topological Defects in Standard Cosmology}

Before discussing what kinds of spectra of gravitational waves the
alternative cosmological scenarios mentioned in Section 3 lead to,
it is important to point out that even in Standard Cosmology
there are processes that lead to a scale-invariant spectrum of
gravitational waves. Specifically, I am referring to gravitational
waves produced by a scaling distribution of topological defects
such as cosmic strings \cite{CSrevs}. 

Cosmic strings are linear topological defects predicted to form
in a wide range of phase transitions of matter in the early
universe. A subset of grand unified field theories leads to
cosmic strings, in particular in many supergravity models \cite{Rachel}.
Cosmic strings also form at the end of inflation in a wide set
of inflationary models which arise motivated by superstring
theory \cite{Tye}. Cosmic strings are characterized by a mass
per unit length $\mu$ (which is conventionally expressed in
terms of the dimensionless number $G \mu$, $G$ being
Newton's gravitational constant). The current upper
bound on $\mu$ is about $G \mu < 2 \times 10^{-7}$ 
from the precision data on the angular power spectrum of CMB
fluctuations in the region of the first acoustic peak \cite{limit}.

In models which admit topologically stable cosmic strings, a
network of such strings will inevitably form in the early
universe (see \cite{CSrevs} for detailed discussions and references
to the original literature). It can be argued analytically and confirmed numerically
that the string distribution will rapidly approach a ``scaling solution''
characterized by a network of infinite strings with a typical
curvature radius which at all times $t$ is of the order the Hubble radius $t$, and
a distribution of string loops with radii $R < \alpha t$ which looks
the same at all times if distances are scaled to the Hubble radius. In
the above, $\alpha$ is a number smaller than $1$ which needs to be determined
by numerical simulations. The loops are formed via intersections
of segments of the infinite string network. They oscillate and
gradually decay via gravitational radiation. 

Due to the fact that the distribution of strings is scaling,
a scale-invariant spectrum of gravitational waves emerges. In
turn, this produces a tensor contribution to the microwave
anisotropies, as first calculated in \cite{BAT} (see also
\cite{others3} for other studies of gravitational radiation
from a network of cosmic strings).

Based on the scaling solution of the network of infinite
strings, one can calculate the distribution of cosmic string
loop sizes. The number density $n(R, t)$ in space of
loops per $R$ interval at time $t$ (i.e. $n(R, t) dR$ is the number density
in space of loops with radii in the interval $[R, R + dR]$),
then for times $t \gg t_{eq}$ (where $T_{eq}$ is the time
of equal matter and radiation) the distribution is given by \cite{TB}:
\bea
n(R, t) \, &=& \, \nu R^{-2} t^{-2} \,\,\, R \gg t_{eq} \nonumber \\
n(R, t) \, &=& \, \nu R^{-5/2} t_{eq}^{1/2} t^{-2} \,\,\, 
\gamma G \mu t < R \ll t_{eq} \, .
\eea
For smaller values of $R$, $n(R, t)$ is constant. The
constant $\gamma$ determines the overall strength of
gravitational radiation, and $\nu$ is a coefficient
which depends on the number per Hubble volume of
infinite string segments in the scaling solution.

Based on this scaling distribution of string loops, the
energy density distribution in gravitational waves can
be calculated, and the resulting CMB temperature
anisotropies can be inferred \cite{BAT}. The 
CMB anisotropy spectrum is scale invariant on large
angular scales and has a r.m.s. amplitude which is
given on angular scales $\chi$ larger than about two
degrees by \cite{BAT}
\bea
< \bigl( \frac{\delta T}{T} \bigr)^2(\chi) >^{1/2} \, 
&\sim& \, [\frac{144}{5} \pi \beta \gamma \nu ]^{1/2} \nonumber \\
&\sim& \, 3 \times 10^{-6} \bigl( \frac{G \mu}{2 \times 10^{-7}} \bigr)
\eea
where in the final step we have inserted typical values
$\beta = 9$, $\nu = 10^{-2}$ and $\gamma = 5$ which follow
from cosmic string network evolution simulations \cite{AT}
\footnote{Taking the numbers from the more updated simulations
of \cite{recent} might change the overall coefficient by
an order of magnitude - reliable statements are impossible
to make since there is still a lot of uncertainty about the
details of the cosmic string loop distribution during the
scaling regime.}. In the above, $\beta$ is a constant
of the order $2 \pi$ which gives the length of a typical
string loop as a multiple of the mean loop radius.

The above result for the amplitude of the tensor mode should be
compared with the observed amplitude of the scalar mode
which is
\be
< \bigl( \frac{\delta T}{T} \bigr)^2(\chi) >^{1/2} \,
\sim \, 10^{-5} \, ,
\ee
on large angular scales. Thus, the predicted tensor to
scalar ratio $r$ is
\be
r \, \sim \,  3 \times 10^{-1} \bigl( \frac{G \mu}{2 \times 10^{-7}} \bigr)\, .
\ee
For strings with tension close to the current observational bound
this is a value of $r$ greater than what most inflationary models
predict.

Although cosmic strings predict a spectrum of gravitational
waves which is scale-invariant on large angular scales, the
model predicts specific non-Gaussian signatures in position
space. In CMB temperature maps the signal is a distribution of
edges across which the temperature jumps \cite{KS}. These
edges are due to strings present between the time $t_{rec}$ of
last scattering and the present time $t_0$. String segments
created at time $t$ lead to discontinuity lines in CMB maps
of comoving scale $s_c$ given by
\be \label{scale}
s_c \, \sim \, t \bigl( z(t) + 1 \bigr) \, ,
\ee
where $z(t)$ is the cosmological redshift at time $t$. These
edges can be searched for using edge detection algorithms
(see e.g. \cite{Amsel, Stewart, Rebecca} for recent studies).

As pointed out in \cite{Holder}, wakes behind moving strings present
between the time $t_{eq}$ of equal matter radiation and the 
present time lead to a specific polarization signal. Most importantly,
the extra scattering which photons passing though a wake at time 
${\tilde t}$ experience leads to rectangular regions in the sky (with 
comoving scale given by (\ref{scale})) of extra polarization. 
When averaged over all strings, there is an equal distribution of
E-mode and B-mode polarization. This example shows that
the discovery of cosmological B-mode polarization is not
necessarily due to primordial gravitational waves.

Since the polarization signal of cosmic strings also has 
special non-Gaussian features, it is important to search
for them in position space through the use of statistics
such as the edge detection algorithms mentioned above.

As initially pointed out in \cite{Krauss}, phase transitions
in Standard Cosmology can lead to a scale-invariant spectrum
of gravitational waves, even if they do not produce defects.
This mechanism is further discussed in \cite{Krauss2}. As
pointed out in these works, a way to potentially differentiate
between inflation and phase transitions as the source of
gravitational waves is the absence of characteristic acoustic
oscillations in the angular power spectrum in the second case.

After this discussion of gravitational waves and direct
B-mode polarization produced by a source which might
be present in Standard Cosmology,
let us now turn to the predictions for the spectrum of the
stochastic background of gravitational radiation from
the alternative cosmological models introduced earlier.

\subsection{Gravitational Waves from String Gas Cosmology}

We have seen in Section 3 that String Gas Cosmology leads
to an almost scale-invariant spectrum of cosmological
perturbations. What is important for this 
Note is that string gas cosmology also leads to an almost
scale-invariant spectrum of primordial gravitational 
waves \cite{BNPV2}. The slope of the spectrum is predicted
to be slightly blue, and not red as inflationary universe models
predict \footnote{The reason for the blue tilt is easy to
understand physically \cite{BNPV2}: the amplitude of the
gravitational waves is given by the anisotropic pressure
fluctuations. Their amplitude, in turn, is proportional to
the background pressure. The background pressure tends
to zero as we go deeper back into the emergent phase. Hence,
long wavelength fluctuations which exit the Hubble radius
earlier (see Figure 4) obtain less gravitational wave power
than short wavelength modes.} The amplitude of the
gravitational wave spectrum and the resulting tensor to 
scalar ratio $r$ can also be calculated
from first principles \cite{BNPV2}. The value of $r$ is related by a
consistency relation to the magnitude of the (red) tilt
$n_s$ of the spectrum of cosmological perturbations:
\be
r \, \sim \, |n_s - 1| \, .
\ee

Comparing this result to what is obtained in inflationary
models, we see that in both scenarios, the typical value
of $r$ is substantially smaller than $r = 1$. Which
scenario gives rise to a larger amplitude depends very
much on the special realization of the scenario.

The key difference between the predictions of inflation
and string gas cosmology relates to the index
of the spectrum. Whereas inflationary models always
yield a red spectrum, string gas cosmology generically
produces a blue shift (which makes it easier to detect
on shorter wavelengths - see \cite{Stewart2} for  some
current bounds). Based on this difference, a key
challenge for experimentalists will be not only to
detect the gravitational wave spectrum through
B-mode polarization, but also to measure the spectral
slope. If the results indicate a blue spectrum, the
entire inflationary paradigm of structure formation
will have been ruled out, and the experiments would
have detected an effect which was first predicted from
superstring theory \cite{BNPV2}.

\subsection{Gravitational Waves from a Matter Bounce}

In the case of the matter bounce scenario, the evolution
of the spectrum of gravitational waves and of cosmological
perturbations is identical on super--Hubble scales during
the contracting phase. If the bounce phase is short,
the evolution in this phase will not lead to a change in the amplitude 
of the fluctuations (neither of cosmological perturbations nor
of gravitational waves). Thus, rather generically a matter bounce
scenario will lead to a scale-invariant spectrum of gravitational
waves with a large amplitude ( $r$ of the order $1$). Taking into
account the presence of entropy modes will lead to a reduction
in the value of $r$ \cite{Cai}, but nevertheless the predicted
amplitude of the gravitational wave spectrum is generically
larger than what is predicted in inflationary models.

The spectrum of fluctuations produced by the matter bounce
can be distinguished from that generated by simple single field inflationary
models in terms of the induced non-Gaussianities. Specifically,
the matter bounce predicts a special shape \cite{BounceNG}
of the bispectrum of cosmological fluctuations. The difference
in shape arises in the following way: the general expression
for the bispectrum is a function of the curvature fluctuation variable
$\zeta$ and its time derivative ${\dot \zeta}$ \cite{Malda}. In an expanding
inflationary universe, $\zeta$ is constant on super-Hubble scales
in the absence of entropy fluctuations, whereas in a contracting universe
the dominant mode of $\zeta$ is increasing. Since the terms involving
${\dot \zeta}$ lead to a diferent shape of the bispectrum than that generated
by the other terms, a matter bounce can in principle be distinguished
from inflationary cosmology by measuring the shape of the bispectrum.

\subsection{No Gravitational Waves in the Ekpyrotic Scenario}

In the case of the Ekpyrotic scenario, the equation of motion
of the cosmological perturbations depends on the potential
of the scalar field, whereas that of the gravitational waves
does not. 

If we start out with vacuum fluctuations for both cosmological
fluctuation and gravitational wave modes on sub-Hubble scales in
the contracting phase, then, unlike in the Matter Bounce scenario,
the increase of the amplitude of the gravitational wave modes on
super-Hubble scales is insufficient to turn the initial vacuum
spectrum into a scale-invariant one. The predicted spectrum of
gravitational waves is very blue and hence primordial 
gravitational waves are negligible on scales of cosmological interest
today. In contrast, the cosmological perturbation modes couple to
the potential and this allows them to attain a scale-invariant
spectrum.

\section{Conclusions}

In this Note I have shown that there are many sources of
gravitational waves which could lead to a roughly scale-invariant
spectrum on cosmological scales. Thus, the detection of 
relic gravitational radiation via B-mode polarization will NOT
prove inflation. Since several of the mechanisms described
here predict a spectral amplitude which is larger than that
generated in the simplest inflationary models, one may argue
that it is more likely that a positive signal will be due
to a source different from inflation. Luckily, the different
sources of gravitational waves - all of them giving a
roughly scale-invariant spectrum - lead to specific predictions
with which they can be distinguished. String gas cosmology
produces a slight blue tilt of the spectrum, whereas inflationary
cosmology always produces a slight red tilt. Cosmic strings lead
to specific non-Gaussian signatures which can be identified in
position space (see e.g. \cite{Amsel,Rebecca} for studies of how
to identify the cosmic string signal in CMB temperature maps).
The matter bounce scenario leads to a particular shape of the
bispectrum \cite{BounceNG}. 

It must also be kept in mind that gravitational radiation is
not the only way to generate primordial B-mode polarization.
Once again, cosmic strings formed in a phase transition during
the early Standard Cosmology phase of the evolution of the universe
will generate primordial polarization which is statistically
equally distributed between E-mode and B-mode \cite{Holder}. This
mechanism produces edges in CMB polarization maps - a signal 
which is easy to look for with position space based algorithms.

The message which this Note is supposed to convey is the following:
the search for B-mode polarization is an extremely interesting
field, much more interesting than if the only source of such
primordial polarization were gravitational waves from inflation.
It will be very important to carefully analyze the data without
the prejudice that inflation is the only source of a signal. Otherwise,
the possible existence of cosmic strings could be missed.
More strikingly, if B-mode polarization is found and is shown
to be due to gravitational waves, then if the spectrum is slightly
blue one would have falsified the inflationary paradigm.

\begin{acknowledgments}

This work is supported by an
NSERC Discovery Grant, by the Canada Research Chairs program and
by a Killam Research Fellowship. I thank Gil Holder, Matt Dobbs 
and Xingang Chen for comments on the draft.

\end{acknowledgments}

\end{document}